\documentclass[12pt,preprint]{aastex}

\newcommand       \Angstrom     {\,{\rm \AA}}

\newcommand       \cm           {\,{\rm cm}}

\newcommand   \g        {\,{\rm g}}

\newcommand       \mum          {\,{\rm \mu m}}

\newcommand       \simali       {\sim\,}
\newcommand   \magni       {\,{\rm mag}}
\newcommand   \amin        {\,{a_{\rm min}}}
\newcommand   \amax        {\,{a_{\rm max}}}

\shorttitle{Dust Extinction of GRB Host Galaxies}

\begin{document}
\title{
Determining the Dust Extinction of Gamma-ray Burst Host Galaxies: A
Direct Method Based on Optical and X-ray Photometry
     }
\author{Y. Li\altaffilmark{1},
        Aigen Li\altaffilmark{1,2},
        and D.M. Wei\altaffilmark{1,3}}

\altaffiltext{1}{Purple Mountain Observatory, Chinese Academy of
                 Sciences, Nanjing 210008, China; {\sf yli@pmo.ac.cn}}
\altaffiltext{2}{Department of Physics and Astronomy, University of
                 Missouri, Columbia, MO 65211; {\sf lia@missouri.edu}}
\altaffiltext{3}{Joint Center for Particle Nuclear Physics and
                 Cosmology of Purple Mountain Observatory --
                 Nanjing University, Nanjing 210008, China;
                 {\sf dmwei@pmo.ac.cn}}

\begin{abstract}
The dust extinction of gamma-ray bursts (GRBs) host galaxies,
containing important clues to the nature of GRB progenitors and
crucial for dereddening, is still poorly known.
Here we propose a straightforward method to determine the extinction
of GRB host galaxies by comparing the observed optical spectra to
the intrinsic ones extrapolated from the X-ray spectra.
The rationale for this method is from the standard fireball model:
if the optical flux decay index equals to that of the X-ray flux,
then there is no break frequency between the optical and X-ray
bands, therefore we can derive the intrinsic optical flux from the
X-ray spectra.
We apply this method to three GRBs of which the optical and X-ray
fluxes have the same decay indices and another one with inferred
cooling break frequency, and obtain the rest-frame extinction curves
of their host galaxies.
The derived extinction curves are gray and do not resemble any
extinction curves of local galaxies (e.g. the Milk Way, the
Small/Large Magellanic Clouds, or nearby starburst galaxies). The
amount of extinction is rather large (with visual extinction
$A_V$\,$\sim$\,1.6--3.4$\magni$).
We model the derived extinction curves in terms of the
silicate-graphite interstellar grain model. As expected from the
``gray'' nature of the derived extinction curve, the dust size
distribution is skewed to large grains.
We determine, for the first time, the local dust-to-gas ratios of
GRB host galaxies using the model-derived dust parameters and the
hydrogen column densities determined from X-ray absorptions.
\end{abstract}


\section{Introduction}
It is widely acknowledged that the long-duration gamma-ray burst
(GRB) is associated with the collapse of a massive star (Woosley
1993). Observational evidence supporting this collapsar model
includes the underlying supernova components in the afterglow of
many GRBs (Zeh et al.\ 2004) and the observed location of GRBs in
star-forming galaxies and active star-forming regions within their
host galaxies (Paczy\'{n}ski 1998, Fruchter et al.\ 2006). In this
scenario, GRBs are born and explode inside dense, dusty
environments. The huge $\gamma$-ray energy emission of GRBs is
almost unaffected by absorptions, allowing them to be detected up to
rather high redshifts (e.g. see Tagliaferri et al.\ 2005).
Therefore, to study the dust and gas properties in the surrounding
vicinity of GRBs is of great significance in understanding the
interstellar medium (ISM) of star-forming galaxies throughout cosmic
history. In addition, an accurate apprehension of the dust and gas
immediate surrounding GRBs can also help (1) to reveal the nature of
so-called ``dark bursts''
    (i.e., whether the non-detection of
     some optical afterglow is due to
    dust extinction
    or the afterglow is intrinsically dark;
    see Lazzati et al.\ 2002 and references therein),
(2) to detect the dust evolution with cosmic time, and (3) to
correct for the extinction of optical emission
    in GRB afterglow analysis.

The dust extinction of GRB host galaxies is traditionally modeled
using either the Milky Way (MW), the Large Magellanic Cloud (LMC),
the Small Magellanic Cloud (SMC), or other presumed extinction
curves (e.g. see Stratta et al.\ 2004; Kann et al.\ 2006; Schady et
al.\ 2007; Starling et al.\ 2007; Tagliaferri et al.\ 2007).
Recently, Chen et al.\ (2006) made the first effort to determine the
extinction curves for GRB host galaxies without a priori assumption
of the extinction law. The derived extinction curves differ from any
known extinction laws of the Milky Way and external galaxies,
challenging the traditional method commonly used in determining the
extinction curves of GRB host galaxies.

In this work we propose a novel, straightforward method to
determine the extinction of GRB host galaxies by comparing the
observed optical spectra to the intrinsic ones extrapolated from the
X-ray spectra. That such an analysis is possible follows from the
standard fireball model. Based on the multi-wavelength afterglow
photometry (including both the X-ray and optical data), we obtain
the extinction curves of four selected bursts. We then model the
size distribution and composition of the dust with the
silicate-graphite interstellar grain model and obtain the
dust-to-gas ratios in the local environment of GRBs.

\section{Method}
The standard fireball model (Sari et al.\ 1998), which has been
successful in explaining the overall properties of GRB afterglow
(M\'{e}sz\'{a}aros \& Rees 1997), predicts that the afterglow
emission is produced by synchrotron radiation of electrons
accelerated by the forward shock. In this model, with typical
parameters, the optical to X-ray spectra can be described by a
broken power law with indices $\beta\,=\,\left(p-1\right)/2$ for
$\nu<\nu_c$ or $\beta\,=\,p/2$ for $\nu>\nu_c$, where $\nu_c$ is the
cooling frequency and $p$ is the electron energy distribution index.
In most cases, the cooling break position is hard to determine. If
the decay indices $\alpha$ of X-ray and optical bands are different,
the cooling frequency lies between them, making it difficult to
calculate the intrinsic optical flux from X-ray data. However, if
the decay indices $\alpha$ of X-ray and optical bands are the same,
then the optical and X-ray should lie on the same spectral segment,
rendering it possible to calculate the intrinsic flux density in any
optical band from
$F_\lambda=F_X\left(\lambda/\lambda_X\right)^{\beta-2}$, where
$\beta$ is the X-ray afterglow spectral index that we get from
fitting the X-ray spectrum, and $F_X$ is the X-ray flux density.
After corrected for the Galactic extinction using the reddening maps
of Schlegel et al.\ (1998), the observed spectral energy
distribution (SED) of GRB at redshift $z$ can be described as
$F_{\lambda\left(1+z\right)}=F_\lambda\exp\left(-A_\lambda/1.086\right)$.
Therefore, the extinction of the GRB host galaxy can be given by
\begin{equation}
A_\lambda\,=\,1.086\,\ln\,\frac{F_X\left(\lambda/\lambda_X\right)^{\beta-2}}{F_{\lambda\left(1+z\right)}}.
\end{equation}
With $A_V$ interpolated, we can obtain the extinction curves
(normalized to $V$ band) of the GRB host galaxies.

We then fit the derived extinction curve $A_\lambda/A_V$ with the
standard silicate-graphite interstellar dust model which has
successfully reproduced the extinction and IR emission of the MW
galaxy, SMC and LMC (Weingartner \& Draine 2001; Li \& Draine 2001,
2002). The grain size distribution for both silicate and graphite is
modeled with: $dn=N\left(a\right)da\varpropto
a^{-\eta}\,\exp(-a/a_c)\,da$, where $a$ is the grain radius (assumed
spherical), ranging from $\amin$\,=\,0.005$\mum$ to
$\amax$\,=\,2.5$\mum$, $a_c$ is the cut-off size. Note that it is
assumed that both silicate dust and graphitic dust have the same
size distribution. Let $f_{\rm gra}$ be the number fraction of
graphitic dust, the mass fraction of graphitic dust is $f_{\rm
gra}^{\prime} = f_{\rm gra}\rho_{\rm gra}/\left[f_{\rm gra}\rho_{\rm
gra}+\left(1-f_{\rm gra}\right)\,\rho_{\rm sil}\right]$, where
$\rho_{\rm sil}\approx 3.5\g\cm^{-3}$ is the mass density  of
silicate material and $\rho_{\rm gra}\approx 2.24\g\cm^{-3}$ is that
of graphite.

With the fitted dust parameters, we can estimate the dust-to-gas
ratio in each of the GRB host galaxies:
\begin{equation}
\frac{m_{\rm dust}}{m_{\rm gas}} =\frac{M_{\rm gra}+M_{\rm
sil}}{1.4\,N_{\rm H}\,\mu_{\rm H}}  ,
\end{equation}
where $N_{\rm H}$ is the hydrogen column density in the host galaxy;
$\mu_{\rm H}$ is the atomic weight of H; the factor ``1.4'' accounts
for helium; $M_{\rm gra}$ and $M_{\rm sil}$ are the column mass
density of graphite and silicate material, respectively:
\begin{equation}
M_{\rm gra}=N_d\,\int_{\amin}^{\amax}
            \frac{4}{3}\pi\,a^{3}
            N\left(a\right)\rho_{\rm gra}f_{\rm gra}\,da ;
\end{equation}
\begin{equation}
M_{\rm sil}=N_d\,\int_{\amin}^{\amax}
            \frac{4}{3}\pi\,a^{3}
            N\left(a\right)\rho_{\rm sil}\left(1-f_{\rm
            gra}\right)\,da .
\end{equation}
$N\left(a\right)$ is the normalized dust size distribution;
The dust column density $N_d$ can be derived from
\begin{equation}
A_{\lambda}=1.086\int_{\amin}^{\amax}
            N\left(a\right)\pi\,a^2\,[f_{\rm gra}\,Q_{\rm ext, \rm gra}(a,\lambda)+\left(1-f_{\rm gra}\right)\,
            Q_{\rm ext, \rm sil}(a,\lambda)]\,da\,N_d ,
\end{equation}
where $Q_{\rm ext, \rm gra}(a,\lambda)$ and $Q_{\rm ext, \rm
sil}(a,\lambda)$ is the extinction efficiency of dust of radius $a$
at wavelength $\lambda$ for graphite and silicate material,
respectively.

\section{Data}
We select four GRBs that have both optical and X-ray observations.
Photometric data are taken from literature (see Tabled 1,2). The
optical to X-ray spectra are extracted when the afterglow
light-curve are in a steady power-law state (e.g. see Panaitescu \&
Kumar 2001, Fan \& Piran 2006 for detailed analysis) to avoid
complex phases (i.e. X-ray flares or re-brightening when the optical
and X-ray emission are probably due to different components [Zhang
et al.\ 2006]; see Fig.\,1). For GRB 020405, GRB 030227 and GRB
060729, we adopt the spectra obtained when the cooling frequency
$\nu_c$ falls below the optical band, indicating an intrinsic single
power law spectrum through optical and X-ray bands as discussed
above. The decay indices $\alpha$ are all taken from literature
except for GRB 060729 (around 0.35 day during plateau phase; for
which $\alpha$ is not available in literature) we derive by fitting
the afterglow light curve between 0.2--0.6\,day. For GRB 061126, the
decay indices of X-ray and optical bands are different, indicating a
break frequency lying between them. At $\sim30\,{\rm ks}$, the R
band afterglow shows a break (see Fig.\,1d), which can be
interpreted as the spectral break passing through the R
band,\footnote{We note, however, that this burst, like many other
Swift bursts, does not obey the closure relation in the standard
afterglow model (Perley et al.\ 2007), which adds uncertainties to
our analysis. But the uncertainty of break frequency does not appear
to affect the shape of the derived extinction curve -- as can be
seen in Figure 2, the extinction curve for the other three bursts
remains gray even if we ignore the GRB 061126 data.} allowing us to
calculate the intrinsic optical flux from
$F_\nu/F_X=(\nu/\nu_c)^{-\beta+0.5}(\nu_c/\nu_X)^{-\beta}$.

\section{Results}
We present in Table 3 the derived $A_\lambda$ of the GRB host
galaxies at every observed optical band and in Table 4 the $V$-band
extinction versus hydrogen column density ($A_V/N_{\rm H}$) and the
dust-to-gas ratio. The errors of the X-ray spectrum can bring about
uncertainties on the extrapolated optical fluxes and thus on the
derived $A_\lambda$. We estimate the errors of $A_\lambda$ from
$\beta$ (the X-ray spectral index) through Eq.1. Larger errors of
the X-ray spectrum result in larger uncertainties in $A_\lambda$
(e.g. see Table 3 and Fig.\,2, GRB 020405). Most noticeably, the
derived extinction curves of the four bursts are rather ``gray''
(see Fig.\,2). Since all these extinction curves have very similar
slope, we put all the extinction data of the four bursts together to
fit them to the silicate-graphite grain model. The best fit
parameters are $\eta\approx 2.60$, $a_c\approx 2.0\mum$, and $f_{\rm
gra}\approx 0$, with $\chi^2 \equiv \sum_{\rm all\,\lambda}\sum_{\rm
all\,GRBs}\left[\left(A_\lambda/A_V\right)_{\rm mod}
 - \left(A_\lambda/A_V\right)_{\rm obs}\right]^2/\sigma^2
\approx 0.12$ (obtained by summing up all wavebands and all GRBs,
where $\sigma$ is the uncertainty for a given GRB at a given band).
A prominent feature is the considerably small $\eta$ (the canonical
value of $\eta$ is 3.5), indicating a grain size distribution skewed
towards substantially large grains. The main reason for $f_{\rm
gra}\approx 0$ is the absence of the 2175$\Angstrom$ extinction bump
in the derived extinction curves which is generally attributed to
small graphitic grains or PAHs.

\section{Discussion}

Previous works concerning dust extinction of GRB host galaxies
mostly focused on fitting the observed photometry with the intrinsic
power-low spectrum reddened by certain ``standard'' extinction
curves inferred from the Milky Way or nearby galaxies (e.g. see
Starling et al.\ 2007). However, lacking a priori knowledge of the
dust properties in high redshift galaxies harboring GRBs, we have no
reason to assume that they are the same as in local universe (e.g.
see Stratta et al.\ 2007). Chen et al.\ (2006), for the first time,
derived the extinction curves of GRBs without a priori assumption of
the extinction law, but they only used the optical data. In this
work, with carefully selected afterglow data covering X-ray to
optical/near-infrared bands, we obtain the extinction curve of four
GRB host galaxies more directly and precisely, based only on the
standard fireball model.

The ``collapsar'' model predicts GRBs to occur in active
star-forming regions similar to Galactic molecular clouds (Jakobsson
et al.\ 2006) which are heavily enshrouded by dust (Trentham et al.\
2002; Tanvir et al.\ 2004). A recent dust scattering model proposed
to account for the shallow-decay phase in Swift X-ray afterglow also
requires large quantities of dust surrounding the GRBs (Shao \& Dai
2007). Observations supporting the existence of large amount of dust
include the emission and absorption features in some X-ray
afterglows (Antonelli et al.\ 2000; Piro et al.\ 1999), large column
densities of heavy elements revealed by optical spectroscopy studies
(Savaglio \& Fall 2004; Savaglio 2006), and the non-detection at
optical wavelengths for more than half the well-localized GRBs
(Jakobsson et al.\ 2004). In contradiction with these evidence,
traditional SED fitting often finds small extinction, primarily
because the best fit model in most cases is the SMC-type extinction
which, with a steep rise into the far ultraviolet (UV), often
requires a small $A_V$ to fit the spectrum (e.g. see Kann et al.\
2006; Schady et al.\ 2007; Tagliaferri et al.\ 2007). Our work,
showing considerably large $A_V$ compared to that fitted with
traditional method, is more consistent with theoretical prediction
and observations. In addition, Rol et al.\ (2007) found that for GRB
051022 a lower limit of $A_V\approx 4.4\magni$ was needed, which
implies that at least in some GRBs the extinction $A_V$ is rather
large.

The extinction curve derived in our work is flat, almost independent
of wavelength, and is even ``grayer'' than the gray type of
extinction curve obtained by Chen et al.\ (2006), similar to the
Calzetti et al.\ (1994) law suitable for local starburst galaxies.
This result is in good agreement with other works fitting the SEDs
of these bursts (e.g. Stratta et al.\ 2005; A. Li et al.\ 2007, in
preparation). In particular, Perley et al.\ (2007) found that for
GRB 061126 the extinction curve is gray. Gray extinction has also
been observed in Galactic dense clouds (Cardelli et al.\ 1988) and
in the circumnuclear region of some AGNs (see Li 2007 for a review).
Gray extinction is produced by a dust distribution biased towards
large grains (see \S4), which may form from (1) grain coagulation
naturally expected
    in the dense circumstance near GRBs
    (Maiolino et al.\ 2001a,b),
(2) the biased evaporation of smaller grains due to
    the intense X-ray and UV radiation up to $\sim$\,20 parsecs
    from the GRB (Waxman \& Draine 2000,
    Fruchter et al.\ 2001, Savaglio et al.\ 2003), and
(3) preferential destruction of small grains by high energy ions
    in fast shocks (Jones 2004).
Perna et al.\ (2003) computed the extinction curve that is obtained
if standard Galactic dust is exposed to a GRB lasting more than a
few tens of seconds (three of the four bursts in our sample meet
this requirement, see $T_{90}$ in Table 1) and found that the
extinction curve can be very flat, chiming with our result. We favor
the grain growth hypothesis since the preferential destruction of
small grains only occurs in the immediate GRB environment
($\sim$10--20 pc from the burst).

It has long been proposed that GRB afterglow radiation, as well as
the prompt emission, can destroy dust grains and cause $A_V$ to
decrease with time (e.g. see Vreeswijk 1999). We test this effect
for GRB 060729, which is exceptionally bright in X-rays as well as
at UV/optical wavelengths showing an unusually long unanimous
plateau phase ($\thicksim 1$\,day). We derive $A_V\approx
1.70\pm0.20\magni$ at $t=0.35$\,day (in the plateau phase) and
$A_V\approx 1.59\pm0.20\magni$ at $t=4.6$\,days (in normal decay
phase) respectively (see Tables 1, 4), showing no significant dust
destruction during this time. Detailed studies of dust destruction
by GRBs will be presented in a forthcoming paper (Z. Jin et al.\ in
preparation).

In accordance with previous works, we find that the average
$A_V/N_{\rm H}$ is
smaller than that in the Milky Way,\footnote{%
  {An exception to this is GRB 060729, for which
  the derived dust-to-gas ratio is $\simali$10 times
  higher than that of the Milky Way (see Table 4).
  We note that, in estimating the dust-to-gas ratios,
  one uncertainty is the equivalent hydrogen column
  densities $N_{\rm H}$ which is derived from the X-ray data
  by assuming a solar metallicity. If we take a metallicity
  of $\thicksim0.1Z_\odot$ for GRB 060729 which is typical
  for GRB host galaxies
  (e.g. Fynbo et al.\ 2006; Stanek et al.\ 2006),
  the resulting dust-to-gas ratio of GRB 060729
  (for which the current value seems to be too high,
   see Table 4) would shrink by roughly an order of magnitude,
   close to that in local galaxies. Besides, photoionization
   of the gas in GRB vicinities by the intense X-ray
   emission can result in a decrease of $N_{\rm H}$
   (Lazzati \& Perna 2002). Richer observations combined
   with more detailed absorption spectroscopic studies will
   help clarify this issue in the near future.}
   }
which is usually ascribed to a lower dust-to-gas ratio in
GRB vicinities (e.g. see Watson et al.\ 2006).
However, there is no obvious reason why the amount
of dust is low in the dense environment surrounding GRBs.
We note that the dust extinction is very sensitive to the dust
size distribution, for larger grains the extinction (on a per unit
mass basis) is low, but the amount of dust may be still high (e.g.
see Li 2007). In fact, based on the model fit dust parameters, the
dust-to-gas ratios for most bursts are larger than that in the Milky
Way. On the other hand, grain growth through coagulation in dense
molecular clouds enshrouding GRBs is expected and this would result
in a dust size distribution biased in favour of large grains, a flat
extinction curve, and a reduced $A_V/N_{\rm H}$.

\acknowledgments We are grateful to the anonymous referee for
helpful comments and suggestions.
We thank S.L. Chen and S.L. Liang for helpful discussion.
D.M.W. is supported by the NSFC grants 10621303 and 10673034,
and the National Basic Research Program of China
(973 Program 2007CB815404). A.L. is supported in part by
a NASA/Swift Theory Program, a NASA/Chandra Theory Program,
and the NSFC Outstanding Oversea Young Scholarship.

\begin{figure}
\begin{center}
\includegraphics[width=0.4\textwidth, bb=17 17 345 244]{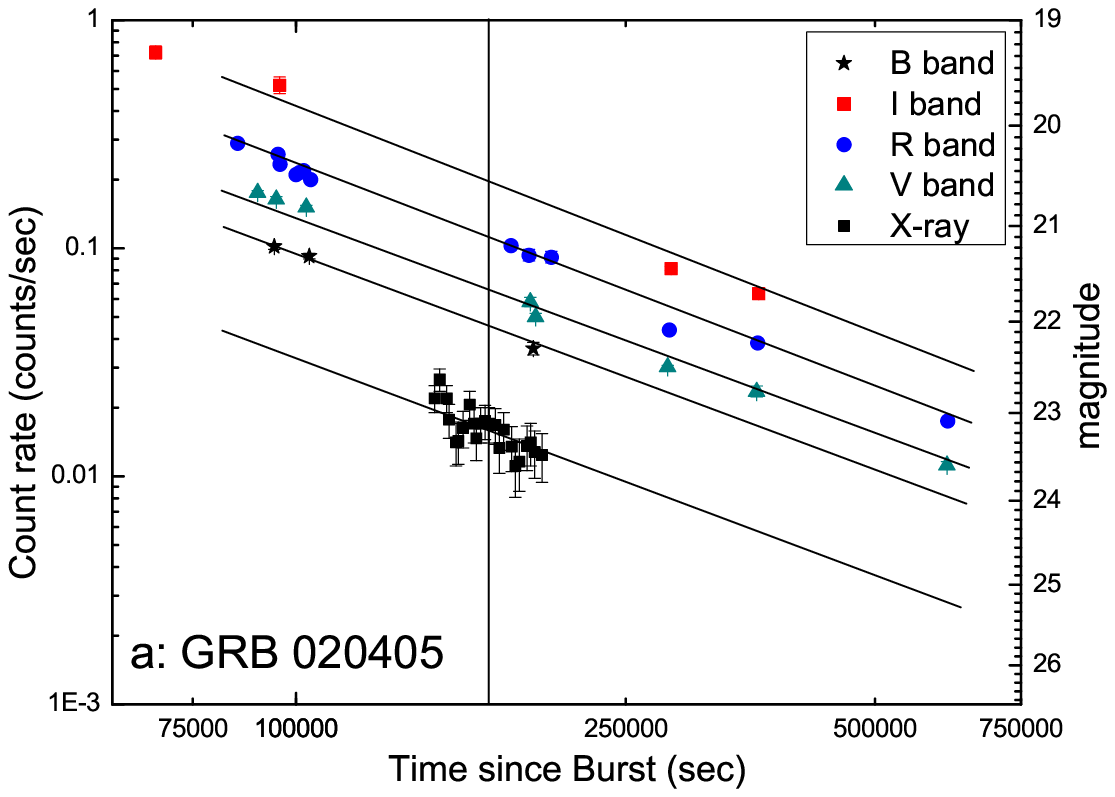}%
\includegraphics[width=0.4\textwidth, bb=17 17 343 260]{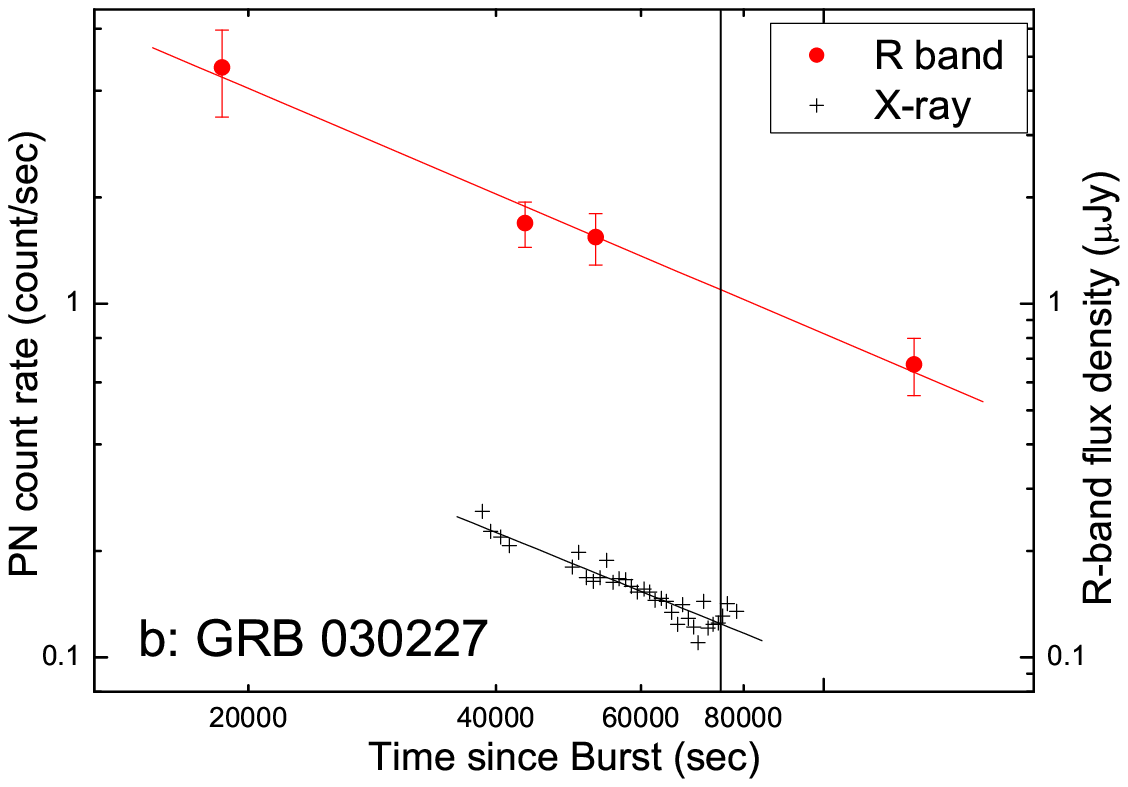}\\
\includegraphics[width=0.4\textwidth, bb=18 17 343 260]{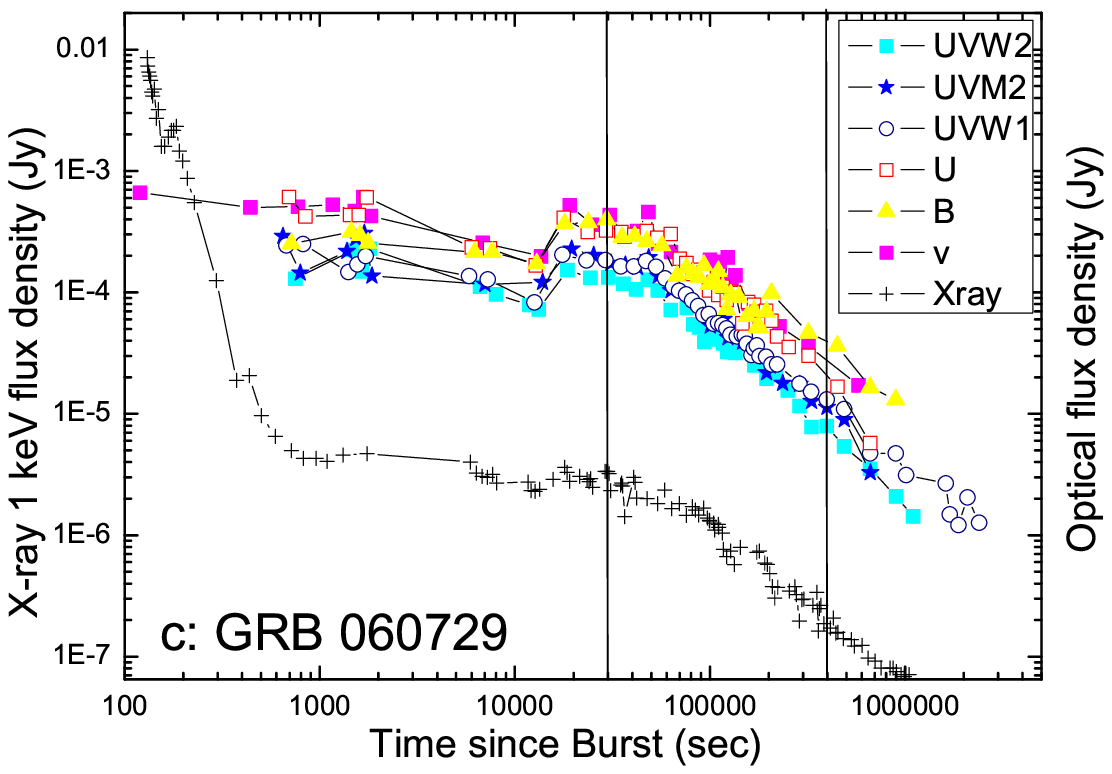}%
\includegraphics[width=0.4\textwidth, bb=17 17 348 264]{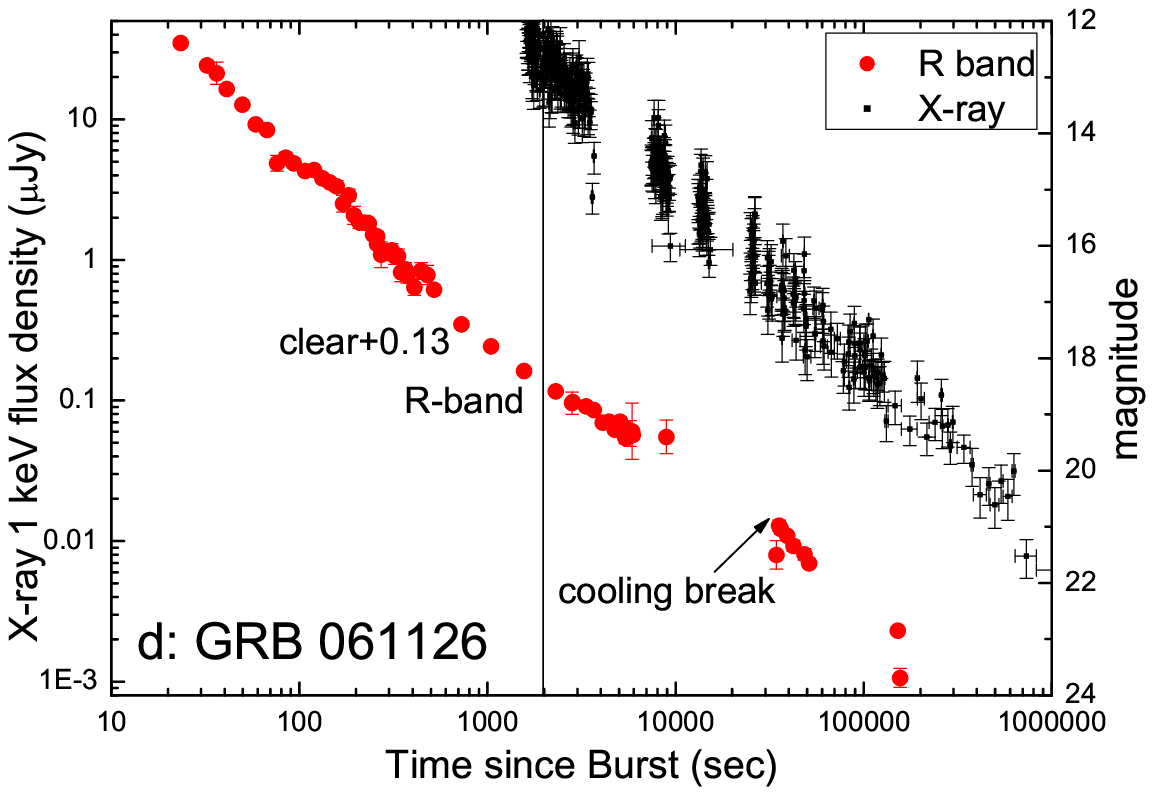}
\end{center}
\caption{\footnotesize\label{fig:result} Light curves of the four
selected bursts. The first three panels (a, b and c) are bursts with
the same decay index in both X-ray and optical bands. The last panel
(d) is for GRB 061126, for which the X-ray and optical bands have
different decay indices and cooling frequency between them. Vertical
lines denote the time when the adopted GRB afterglow spectra were
obtained. }
\end{figure}

\begin{figure}
\begin{center}
\includegraphics[width=0.8\textwidth]{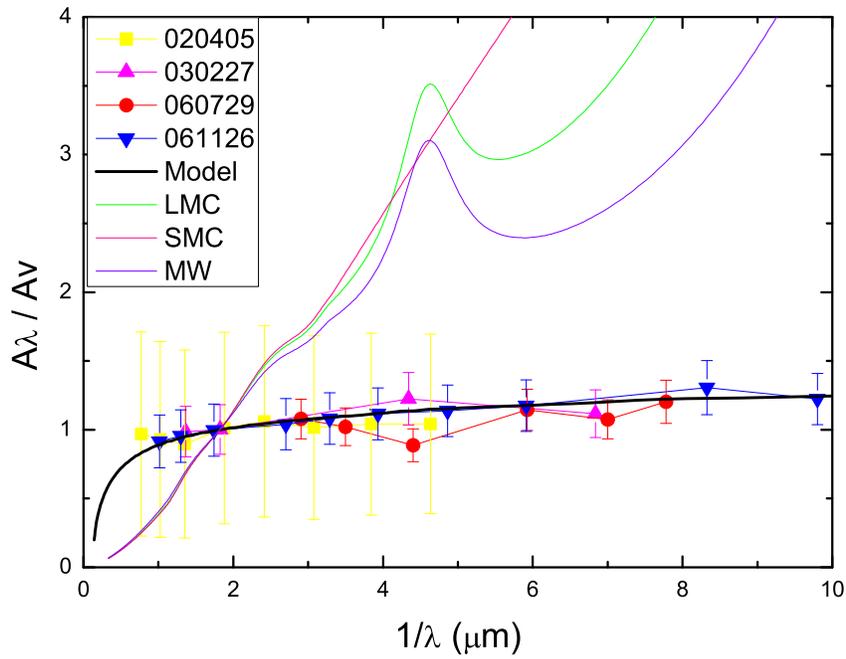}
\end{center}
\caption{\footnotesize\label{fig:result} Rest-frame extinction
curves (normalized to the $V$ band) for 4 GRB host galaxies and the
best-fit extinction curve from the silicate-graphite model (thick
black line). Also shown are the Milky Way, SMC and LMC extinction
curves.}
\end{figure}

\begin{table}
\caption[]{Observational Properties of the four
GRBs\label{tab:tbl1}} {\footnotesize
\begin{tabular}{cccccccccc}
\tableline GRB & $z$ & $\beta$ & $F_X$ ($\mu$Jy) & $t$ (days)& $T_{90}$ (s) & $\alpha_{o}$ & $\alpha_{X}$ & References\\
\tableline\tableline
020405 &0.691   & $1.0\pm0.2$   &0.23  &1.98 & 60  & $1.54\pm0.06$ & $1.97\pm1.10$ & 1, 2 \\
030227 &4       & $0.94\pm0.05$ &0.125 &0.87 & 18  & $0.95\pm0.16$ & $0.97\pm0.07$ & 3, 4\\
060729 &0.54    & $1.06\pm0.01$ &0.2   &4.6  & 115 & $1.27\pm0.10$ & $1.29\pm0.03$ & 5 \\
061126 &1.1588  & $0.5\pm0.07$  &40.5  &0.023& 191 & $0.75\pm0.06$ & $1.31\pm0.01$ & 6 \\
\tableline
060729*$^1$&0.54& $1.06\pm0.01$ &3.46  &0.35 & 115 & $0.26\pm0.07$ & $0.35\pm0.15$ &5\\
\hline
\end{tabular}
\tablecomments{$z$ is the redshift of the burst; $\beta$ is the
intrinsic optical/UV to near-IR spectral index derived from the
standard afterglow model; $F_X$ is the X-ray flux density at 1\,keV;
$t$ is measured from the burst trigger time; $T_{90}$ is the
duration of the burst; $\alpha_{\rm o}$ and $\alpha_{X}$,
respectively the temporal decay index at optical and X-ray bands,
are all taken from literature except GRB 060729, for which we obtain
$\alpha_{\rm o}$ and $\alpha_{X}$ by fitting the the afterglow light
curve between 0.2--0.6\,day.} \tablerefs{(1) Berger et al.\ 2003;
           (2) Stratta et al.\ 2005;
           (3) Castro-Tirado et al.\ 2003;
           (4) Mereghetti et al.\ 2003;
           (5) Dirk Grupe et al.\ 2007;
           (6) Perley et al.\ 2007.}}
\tablenotetext{1}{Data are taken during the plateau phase.
                  See \S5 for discussion}
\end{table}

\begin{table}
\caption[]{Optical/UV to near-IR flux density \label{tab:tbl2}}
{\footnotesize
\begin{tabular}{ccccccccccccccc}
\hline
GRB &\multicolumn{11}{c}{$F_\nu$ ($\mu$Jy)} \\
\cline{2-12}
 &  $UVW2$  &  $UVM2$  &  $UVW1$ & $U$ & $B$ & $V$ & $R$ & $I$  &$J$   &$H$  & $K$\\
\cline{2-12} \tableline\tableline

020405 &\nodata &\nodata &\nodata &5.93    &7.18   &9.5     &10.9    &15.7    &28.6    &34.9    &42.5\\
030227 &\nodata &\nodata &\nodata &\nodata &2.4    &\nodata &2.7     &\nodata &\nodata &10.2    &15  \\
060729 &7.56    &10.18   &11.05   &21.98   & 23    &25.79   &\nodata &\nodata &\nodata &\nodata &\nodata \\
061126 &\nodata &65.12   &54.66   &97.41   &120.18 &143.18  &173.6   &217.97  &309.38  &409.42  &523.16\\
\hline
060729*&124.2   &170.1   &195.3   &313.1   &359.4  &455.1   &\nodata &\nodata &\nodata &\nodata &\nodata \\
\hline
\end{tabular}
\tablecomments{The flux densities $F_\nu$ measured in the observer
frame are taken at the time of the vertical lines reported in
Fig.\,1. All the data have been corrected for Galactic extinction} }
\end{table}

\begin{table}
\caption[]{Derived extinction in the observed wavelength $\lambda$.
           \label{tab:tbl3}} {\tiny
\begin{tabular}{ccccccccccccccccc}
\tableline
GRB  &\multicolumn{11}{c}
{$A_\lambda$} \\
\cline{2-12}
 &  $UVW2$  &  $UVM2$  &  $UVW1$ & $U$  &$B$  & $V$  &$R$ & $I$ &$J$ &$H$ &$K$\\
\cline{2-12} \tableline \tableline
020405 &\nodata      &\nodata       &\nodata      &$2.66\pm1.12$ &$2.65\pm1.16$ &$2.60\pm1.21$ &$2.70\pm1.26$ &$2.58\pm1.31$ &$2.29\pm1.39$ &$2.37\pm1.45$ &$2.47\pm1.51$ \\
030227 &\nodata      &\nodata       &\nodata      &\nodata       &$2.88\pm0.26$ &\nodata       &$3.16\pm0.28$ &\nodata       &\nodata       &$2.58\pm0.33$ &$2.54\pm0.35$ \\
060729 &$1.91\pm0.05$&$1.71\pm0.05$ &$1.82\pm0.05$&$1.41\pm0.05$ &$1.62\pm0.06$ &$1.71\pm0.06$ &\nodata       &\nodata       &\nodata       &\nodata       &\nodata       \\
061126 &\nodata      &$4.09\pm0.03$ &$4.38\pm0.03$&$3.93\pm0.04$ &$3.81\pm0.04$ &$3.74\pm0.04$ &$3.63\pm0.04$ &$3.49\pm0.04$ &$3.34\pm0.05$ &$3.19\pm0.05$ &$3.06\pm0.05$ \\
\hline
060729*&$2.02\pm0.05$&$1.79\pm0.05$ &$1.84\pm0.05$&$1.67\pm0.05$ &$1.78\pm0.05$ &$1.74\pm0.05$ &\nodata       &\nodata       &\nodata       &\nodata       &\nodata       \\
\tableline
\end{tabular}}
\end{table}

\begin{table}
\caption[]{$V$-band extinction and dust-to-gas ratio
           in the rest frame of each burst.
           \label{tab:tbl3}} {\tiny
\begin{tabular}{ccccccccccccccccc}
\tableline GRB & $A_V$(mag) & $N_{\rm H}$$(10^{22}\cm^{-2})$&
$A_V/N_{\rm H}$$(10^{-22}\magni\cm^2)$ & $m_{\rm dust}/m_{\rm
gas}$$(10^{-2})$\\
\tableline\tableline
020405 &$2.50\pm1.17$ &$0.8\pm0.2$ &$3.2\pm1.6$ &$0.99\pm0.52$\\
030227 &$2.57\pm0.32$ &$6.8^{+1.8}_{-3.8}$ &$0.38^{+0.22}_{-0.11}$& $0.12^{+0.07}_{-0.03}$\\
060729 &$1.59\pm0.20$ &$0.076\pm0.003$ & $20.9\pm2.8$ &$6.5\pm0.9$\\
061126 &$3.35\pm0.43$ &$1.1\pm0.3$ &$3.0\pm0.9$ &$0.94\pm0.28$\\
\hline
060729*&$1.70\pm0.20$ &$0.076\pm0.003$ &$22.4\pm2.8$ &$6.9\pm0.9$\\
\hline
\end{tabular}
\tablecomments{$A_V$ is the rest-frame $V$-band extinction of GRB
host galaxies; $N_{\rm H}$ is the rest-frame equivalent column
densities of hydrogen measured from X-ray absorption assuming a
solar metal abundance at the same time when the multi-band spectra
were taken (see \S2)}}
\end{table}
\end{document}